\documentclass[12pt]{article}
\usepackage{epsfig}
\input epsf.sty
\input axodraw.sty
\newcommand{\la}[1]{\label{#1}}
 
\newcommand{\be}{\begin{equation}}
\newcommand{\ee}{\end{equation}}
\newcommand{\ba}{\begin{eqnarray}}
\newcommand{\ea}{\end{eqnarray}}
\newcommand{\bi}{\begin{itemize}}
\newcommand{\ei}{\end{itemize}}
\newcommand{\rmi}[1]{{\mbox{\scriptsize #1}}}
\newcommand{\nr}[1]{(\ref{#1})}

\newcommand{\re}{{\rm Re\,}}
\newcommand{\Hc}{{\rm H.c.\ }}
\newcommand{\im}{\mathop{\rm Im}}
\newcommand{\nn}{\nonumber \\}
\newcommand{\fr}[2]{{\frac{#1}{#2}}}
\newcommand{\msbar}{\overline{\mbox{\rm MS}}}

\renewcommand{\vec}[1]{{\bf #1}}
\newcommand{\bmu}{\bar{\mu}}

\newcommand{\Tint}[1]{{\hbox{$\sum$}\!\!\!\!\!\!\int}_{\!\!\!\!#1}}
\def\ZZ{Z \kern -.53em Z}
 
\newcommand{\eq}{Eq.~}
\newcommand{\eqs}{Eqs.~}
\newcommand{\fig}{Fig.~}

\newcommand{\se}{Sec.~}

\def\lsi{\raise0.3ex\hbox{$<$\kern-0.75em\raise-1.1ex\hbox{$\sim$}}}
\def\gsi{\raise0.3ex\hbox{$>$\kern-0.75em\raise-1.1ex\hbox{$\sim$}}}
\newcommand{\lsim}{\mathop{\lsi}}
\newcommand{\gsim}{\mathop{\gsi}}
\def\apmm{\raise0.3ex\hbox{$\scriptstyle{+}$%
\kern-0.82em\raise-1.1ex\hbox{$\scriptstyle{(-)}$}}}
\newcommand{\pmm}{\mathop{\apmm}}
\makeatletter \@addtoreset{equation}{section} \makeatother
\renewcommand{\theequation}{\arabic{section}.\arabic{equation}}

\begin{document}

\begin{titlepage}
\begin{flushright}
CERN-TH/2000-072\\
hep-ph/0003111
\end{flushright}
\begin{centering}
\vfill

{\bf TWO-LOOP DIMENSIONAL REDUCTION AND EFFECTIVE POTENTIAL 
WITHOUT TEMPERATURE EXPANSIONS}

\vspace{0.8cm}

M. Laine$^{\rm a,b,}$\footnote{mikko.laine@cern.ch} and 
M. Losada$^{\rm c,}$\footnote{malosada@venus.uanarino.edu.co} \\

\vspace{0.3cm}
{\em $^{\rm a}$Theory Division, CERN, CH-1211 Geneva 23,
Switzerland\\}
\vspace{0.3cm}
{\em $^{\rm b}$Dept.\ of Physics,
P.O.Box 9, FIN-00014 Univ.\ of Helsinki, Finland\\}
\vspace{0.3cm}
{\em $^{\rm c}$Centro de Investigaciones, 
Universidad Antonio Nari\~{n}o, \\ 
Cll. 59 No. 37-71, Santa Fe de Bogot\'{a}, Colombia}

\vspace{0.7cm}
{\bf Abstract}

\end{centering}

\vspace{0.3cm}\noindent 
In many extensions of the Standard Model, finite temperature
computations are complicated by a hierarchy of zero temperature 
mass scales, in addition to the usual thermal mass scales. 
We extend the standard thermal resummations to such a situation,  
and discuss the 2-loop computations of the Higgs effective potential, 
and an effective 3d field theory for the electroweak phase transition, 
without carrying out high or low temperature expansions for the 
heavy masses. We also estimate the accuracy of the temperature expansions 
previously used for the MSSM electroweak phase transition in the 
presence of a heavy left-handed stop. We find that the low
temperature limit of dealing with the left-handed stop
is accurate up to surprisingly high temperatures. 

\vfill

\noindent
CERN-TH/2000-072\\
March 2000

\vfill

\end{titlepage}

\section{Introduction}

Electroweak baryogenesis in the Minimal Supersymmetric 
Standard Model (MSSM) is a viable option 
for explaining the matter-antimatter asymmetry observed in the
present Universe, provided that there is a mild hierarchy between 
the right and left-handed stop masses~\cite{bjls}--\cite{cms}. The dominantly 
right-handed stop should be lighter than the top in order to 
make a strong transition, yet the left-handed stop should be rather 
heavy, $\sim 1$ TeV, in order to raise the Higgs mass upper bound 
to $\sim 110$ GeV. Various details of the electroweak phase transition 
in this regime are constantly being investigated~\cite{recent}. 

Here we will  be concerned with the thermodynamics 
of the phase transition. In the perturbative approach, 
this problem is approached by computing the effective 
potential for the Higgs field to some order in the loop 
expansion. In general, such a computation in a weakly coupled 
gauge theory faces two problems: (i) The system has
a hierarchy of mass scales ($2\pi T, gT, g^2 T$), 
which spoils a straightforward
perturbative computation. Historically, this was observed
by finding large ``linear terms'' at 2-loop level~\cite{mes}, 
which were then shown to be absent after
an appropriate resummation~\cite{Dine:1992wr}.  
(ii) At momenta of the order of 
the lowest of the mass scales ($g^2 T$), the system is 
also inherently non-perturbative~\cite{Linde:1980ts}. 

The resummations needed at 2-loop level for dealing with the 
heavy scale $2\pi T$ 
were discussed in detail by Arnold and Espinosa~\cite{ae}. 
However, the problem can also be dealt with in another way, namely
by constructing a sequence of effective field theories by integrating
out, to a given order in perturbation theory,  the scales 
$2\pi T, gT$~\cite{generic,jp}. 
This construction 
is highly accurate in the Standard Model~\cite{cfh,jhep}.
The final theory is three-dimensional (3d), purely bosonic, and  
contains only the momentum scale $g^2T$.  
A perturbative analysis of the 3d theory automatically reproduces 
the results of the resummed 4d effective potential, but the theory 
can also be studied efficiently with 
relatively simple lattice simulations~\cite{nonpert}, 
to account for the non-perturbative part. 

The problem we consider here is the observation that 
the hierarchy of mass scales can be even more severe
in extensions of the Standard Model such as the MSSM. Indeed, 
there one tends to have new mass parameters that are not 
related to the temperature in the same way as $m_H$ is
in the Standard Model, where $m_H\sim gT_c$. 
In particular, as mentioned above, one prefers rather large 
left-handed squark mass parameters, say $m_Q\sim $ 1 TeV. 
Previously, the effects of $m_Q$ have been considered 
(on the 2-loop or non-perturbative level) only 
in the high temperature expansion, 
or in the extreme limit $m_Q\gg 2\pi T$ where the finite 
temperature effects decouple completely. 

Our objective here is to treat in some detail the general 
situation $m_Q\sim 2 \pi T$. First of all, we discuss how
the resummations used previously need to be changed in 
such a situation (\se\ref{linear}). We then show with 
a simple example how the full resummed 2-loop effective
potential could be computed without any temperature expansions
related to~$m_Q$, and how the result can be used for a 2-loop
computation of the mass parameter of an effective 3d field
theory (\se\ref{2loop}). Finally we consider a particular
observable sensitive to $m_Q$, the critical temperature
of the electroweak phase transition, and estimate the  
accuracy of the high and low temperature expansions
employed earlier on (\se\ref{accuracy}). 
We conclude in \se\ref{concl} 
and discuss several possible extensions 
of the computations presented in this paper. 
The expressions used for the 1-loop tadpole and bubble, as well as
2-loop sunset graphs are discussed in the appendices.

\section{Parametric conventions}
\la{param}

In order to be explicit yet concise, 
we illustrate the situation with a simple model
reminiscent of the scalar sector of the MSSM. We take
\ba
{\cal L} & = &   
m_H^2 H^\dagger H + m_U^2 U^*_\alpha U_\alpha + 
m_Q^2 Q^\dagger_\alpha Q_\alpha \nn
&  + &  
h_1^2 H^\dagger H U^*_\alpha U_\alpha + 
h_2^2 H^\dagger Q_\alpha Q^\dagger_\alpha H + 
h_3 \Bigl( A H^\dagger Q_\alpha U_\alpha + \Hc \Bigr)+... \, .
\la{lagr}
\ea
Here $H$ is an SU(2) doublet, $U$ an SU(3) (anti-)triplet,
while $Q$ changes under both groups. We ignore gauge 
interactions for the moment. We assume that 
$h_1\sim h_2\sim h_3\sim g$ are small couplings, 
and $m_H^2,m_U^2 \sim (gT)^2$. 
It is also important to specify the order of 
magnitude of the dimensionful parameter $A$ in \eq\nr{lagr}. 
In this paper we work under the assumption that 
\be
|\hat A|^2 \equiv \frac{|A|^2}{m_Q^2}\sim g^2, \la{AA}
\ee
which simplifies the procedure considerably. 

In the imaginary time formalism, the fields in \eq\nr{lagr}
can be divided into Matsubara modes. We assume that the only
{\em light modes} are the zero Matsubara modes of $H,U$. 
The non-zero Matsubara modes of $H,U$ have effectively 
a mass parameter $\ge (2\pi T)^2$. For the field $Q$, we 
assume that $m_Q$ itself is large, $m_Q\sim 2 \pi T$, so 
that even the zero Matsubara mode is heavy. If
$m_Q\sim gT$, then the zero Matsubara mode of $Q$
is light as well and the procedure is 
the one described in~\cite{old}. If $m_Q\sim 2 \pi T /g$,
on the other hand, $Q$ can be integrated out at $T=0$ 
with exponentially small corrections.  

The issue of resummation can now be formulated as follows. 
Due to the presence of the heavy mass scales, the $n=0$ modes 
of $H,U$ can receive radiative corrections as large as the 
tree-level terms, $\delta m_{H,U}^2 \sim g^2 m_Q^2 + g^2 T^2$. 
Such corrections have to be resummed. In fact, close to 
the phase transition point, the effective mass parameters
$m_{H\rm{eff},U\rmi{eff}}^2$ can be even smaller, of the non-perturbative
magnitude $\sim (g^2 T)^2$. Then resummation has to be extended
to the 2-loop level. Non-zero Matsubara modes, or the field $Q$, 
on the other hand, do not require resummation~\cite{ae}, since the 
mass corrections 
$g^2T^2,g^2 m_Q^2$ are according to our convention small compared 
with the tree-level terms.

This statement can be formulated more precisely as follows. 
Let us write down the effective Lagrangian obtained after 
integrating out all the heavy modes. The light fields being 
the $n=0$ modes of $H,U$, the form of the Lagrangian is  
\be
{\cal L}_\rmi{eff} = 
m_{H\rmi{eff}}^2 H^\dagger H + 
m_{U\rmi{eff}}^2 U^*_\alpha U_\alpha + 
h_{1\rmi{eff}}^2 H^\dagger H U^*_\alpha U_\alpha + ...\, . 
\la{3d}
\ee
Our aim is now to compute expressions of the form\footnote{In a gauge
theory there are also corrections of order $g^3T^2$.} 
\be
m_{H\rmi{eff}}^2 = m_H^2 + c_1 g^2 m_H^2 + c_2 g^2 m_U^2 + 
c_3 g^2 m_Q^2 (1 + c_4 g^2)+
c_5 g^2 T^2 (1 + c_6 g^2), \la{paramform}
\ee
where the $c_i$'s are numerical coefficients. 


\section{Leading order resummation}
\la{linear}

\begin{figure}[t]

\begin{centering}

\parbox[c]{2.4cm}{
\begin{picture}(60,40)(0,0)

\SetWidth{1.5}
\SetScale{0.9}

\CArc(30,20)(15,0,360)
\GCirc(30,35){3}{0}

\Text(30,-10)[]{$P_1$}

\end{picture}}%
\parbox[c]{2.6cm}{
\begin{picture}(70,40)(0,0)

\SetWidth{1.5}
\SetScale{0.9}
\CArc(20,20)(15,0,360)
\CArc(50,20)(15,0,360)

\Text(30,-10)[]{$P_1P_2$}

\end{picture}}%
\parbox[c]{2.4cm}{
\begin{picture}(60,40)(0,0)

\SetWidth{1.5}
\SetScale{0.9}
\CArc(30,20)(15,0,360)
\Line(15,20)(45,20)

\Text(30,-10)[]{$P_1P_2P_3$}

\end{picture}}

\vspace*{1.0cm}

\end{centering}

\caption[a]{The 2-loop graphs considered. The blob means 
a counterterm. Different particles are denoted by $P_i = H, U, Q$.}  
\la{fig:graphs}
\end{figure}
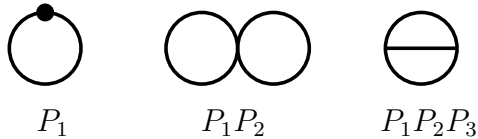

In order to carry out the resummation explicitly, 
let us consider the effective potential of the theory. 
To illustrate the procedure, it is enough to consider only 
the field $H$, keeping the expectation value of $U$ at zero. 
Introducing 
$\langle H\rangle = (0\, , \phi)^T/\sqrt{2}$
changes the mass spectrum of 
the system, $m \to m_\phi = m+\delta_\phi m$. We are 
interested in a certain range of $\phi\sim 0... T$.
Then, in the case of heavy modes, $\delta_\phi m \ll m$, 
and we can expand in $\delta_\phi m$, while in the case of 
light modes we cannot. For the purpose of illustration, 
let us suppress $m$ for the light modes here. 

Then, in the standard thermal case, 
the 1-loop and 2-loop contributions 
to the effective potential behave at small $\phi$ as
\ba
\delta V_\rmi{1-loop} & \sim &
 T^2 (\delta m_\phi)^2 + T (\delta m_\phi)^3 + ... ,
\la{dv1} \\
\delta V_\rmi{2-loop} & \sim  &
g^2 T^3\delta m_\phi+g^2 T^2(\delta m_\phi)^2+...\, . 
\la{dv2}
\ea
The statement of resummation is now that the dominant 2-loop terms, 
the ``linear'' ones $\sim g^2 T^3\delta m_\phi$, arise from a badly 
convergent series which 
can be resummed into a better convergent one~\cite{Dine:1992wr}. 
The way the resummation proceeds is obvious from \eqs\nr{dv1}, \nr{dv2}:
the non-analytic 1-loop and 2-loop terms combine to 
\be
T (\delta m_\phi)^3 + g^2 T^3\delta m_\phi \to 
T (g^2 T^2 +\delta m_\phi^2)^{3/2}.
\ee
This corresponds simply to the corrections of order
$g^2T^2$ in \eq\nr{paramform}.
The extension we make here 
is that when $m_Q\sim 2\pi T$, 
the contribution to be resummed goes from $g^2T^2$
to a non-trivial 
function $g^2 T^2 f(m_Q/T, |\hat A|^2)$.

In order to proceed systematically, we write the mass parameters
related to the light modes as 
\be
m_H^2  = m_{H\rmi{eff}}^2 - \delta_r m_H^2, \quad
m_U^2  = m_{U\rmi{eff}}^2 - \delta_r m_U^2, \la{resum}
\ee
where $m_{H\rmi{eff}}^2,m_{U\rmi{eff}}^2$ appear in the propagators, 
and $\delta_r m_H^2, \delta_r m_U^2$ are treated as interactions. 
Denoting the heavy modes by solid lines and the light modes by dashed 
lines, the graphs
\vspace*{0.4cm}

\hspace*{3cm}%
\parbox[c]{5.4cm}{
\begin{picture}(140,40)(0,0)

\SetWidth{1.5}
\SetScale{0.9}
\DashLine(0,5)(60,5){5}
\CArc(30,20)(15,0,360)

\end{picture}}
\parbox[c]{5.4cm}{
\begin{picture}(140,40)(0,0)

\SetWidth{1.5}
\SetScale{0.9}
\DashLine(0,20)(20,20){5}
\DashLine(50,20)(70,20){5}
\CArc(35,20)(15,180,360)
\CArc(35,20)(15,0,180)

\end{picture}}

\noindent
suggest that 
\ba
\delta_r m_H^2  & = & 3 h_1^2 I_{n\neq 0}(m_U) + 3 h_2^2 I(m_Q) + 
3 h_3^2 |\hat A|^2 \Bigl[ I(m_Q) - I_{n\neq 0}(0)\Bigr], \la{resum0} \\ 
\delta_r m_U^2  & = & 2 h_1^2 I_{n\neq 0}(m_H) + 
2  h_3^2 |\hat A|^2 \Bigl[ I(m_Q) - 
I_{n\neq 0}(0)\Bigr]. \la{resum2}
\ea
Here $I,I_{n\neq 0}$ are tadpole integrals defined
in \eqs\nr{tadpole}, \nr{Ine0}, and we have
made use of $m_H^2,m_U^2\ll m_Q^2$. 
Note that the fact that $|\hat A|^2$ is small, \eq\nr{AA}, 
implies that wave function corrections need not be considered, 
since their effect would be of order 
$\sim h_3^2 |\hat A|^2 m_H^2 \sim g^4 m_H^2$, 
beyond \eq\nr{paramform}. For the same reason, 
we have dropped any $m_H^2,m_U^2$ dependence in 
the terms proportional to $|\hat A|^2$.

In addition to the mass parameters of the scalar fields, resummation 
of course also affects the zero components of the gauge fields. In fact, 
as is well known~\cite{mes,Dine:1992wr}, in the Standard Model
the latter effect is more important for physical observables such as the 
strength of the phase transition, while the former is important 
particularly for the critical temperature. We do not 
discuss infrared dominated observables such as the strength of 
the phase transition, nor gauge fields, to any 
length in this paper, but let us nevertheless note that the 
contributions of $H,U,Q$ to the Debye masses of the SU(2)
and SU(3) fields $A_0,C_0$ are, in the presence of $m_Q\sim 2\pi T$, 
\ba
\delta_r m_{A_0}^2 & = &  g^2 T\frac{d}{dT} 
\Bigl(I_{n\neq 0}(m_H) + 3 I_T(m_Q)  \Bigr), \la{debye1} \\
\delta_r m_{C_0}^2 & = & g_S^2 T\frac{d}{dT} 
\Bigl(I_{n\neq 0}(m_U) + 2 I_T(m_Q)  \Bigr), \la{debye}
\ea
where $g_S$ is the SU(3) gauge coupling
and $I_T$ is defined in \eq\nr{IT}.  In addition
to these terms, the Debye masses of course 
contain the usual gauge and fermion contributions.

In order to now show that the procedure
introduced in \eqs\nr{resum}, \nr{resum0}, \nr{resum2} is 
a consistent one, 
we need to demonstrate that all ``linear terms'' at 2-loop level
cancel, and the remainder is quadratic in $\delta m_\phi$. 
Recalling that we have set the quartic Higgs self-coupling 
to zero (at tree-level) for the purpose of simplicity,  
we get for the shifts in the mass parameters
($Q_{1(2)}$ denote the upper (lower) SU(2) component of $Q$) 
\ba
\delta_\phi m_{H\rmi{eff}}^2 & = & 0, \qquad 
\delta_\phi  m_{U\rmi{eff}}^2 = \fr12 h_1^2 \phi^2 , \\
\delta_\phi m_{Q_1}^2 & = & 0, \qquad
\delta_\phi m_{Q_2}^2 = \fr12 h_2^2 \phi^2.
\ea
Note that due to the assumption $|\hat A|^2 \sim g^2$
we can ignore all corrections involving $\hat A$ here, since
the corresponding 2-loop contributions are at most of
order $\sim h_i^2 h_3^2 |\hat A|^2\sim g^6$. We will denote 
$(m_{H\rmi{eff}}^\phi)^2 = m_{H\rmi{eff}}^2 + 
\delta_\phi m_{H\rmi{eff}}^2$, etc.

Linear terms in the effective potential arise from 
graphs of the types $(H)$, $(U)$, $(HU)$, $(HQ)$, $(HUQ)$ in 
the notation of \fig\ref{fig:graphs}. Denoting by $I_\rmi{3d}$
the 3d tadpole in \eq\nr{I3d} and by $H$ 
the bosonic sunset integral in \eq\nr{sunset}, we obtain
\ba
(H)+(U)  & = &  
-2 \delta_r m_H^2 I_\rmi{3d}(m_{H\rmi{eff}}^\phi)
-3 \delta_r m_U^2 I_\rmi{3d}(m_{U\rmi{eff}}^\phi), \la{Tct} \\
(HU)+(HQ) & = &
6 h_1^2 I(m_{H\rmi{eff}}^\phi)I(m_{U\rmi{eff}}^\phi) + 
3 h_2^2 I(m_{H\rmi{eff}}^\phi) \Bigl[ I(m_{Q_1}^\phi)+ 
I(m_{Q_2}^\phi) \Bigr], \\
(HUQ)  & = & -3 h_3^2 |A|^2 \Bigl[
H(m_{Q1}^\phi,m_{H\rmi{eff}}^\phi,m_{U\rmi{eff}}^\phi) + 
H(m_{Q2}^\phi,m_{H\rmi{eff}}^\phi,m_{U\rmi{eff}}^\phi)\Bigr].
\mbox{\hspace*{1.0cm}}
\ea
We then expand these contributions in $\delta_\phi m$. 
Employing the expansions 
\ba
I(m_Q^\phi) & =  & I(m_Q) - \delta_\phi m_Q^2 D(m_Q) + 
{\cal O}(\delta_\phi m_Q)^4, \\
I(m_\rmi{eff}^\phi) & =  & I_{n\neq 0}(m_\rmi{eff}) + 
I_\rmi{3d}(m_\rmi{eff}^\phi) - 
\delta_\phi m_\rmi{eff}^2 D_{n\neq 0}(m_\rmi{eff}) +
{\cal O}(\delta_\phi m_\rmi{eff})^4, \\
H(m_{Q}^\phi,m_{H\rmi{eff}}^\phi \!\!\!\!\!\! &,& \!\!\!\!\!\! 
m_{U\rmi{eff}}^\phi) \;\; = \;\;   
\frac{1}{m_Q^2}\Bigl[ \Bigl( 
I_\rmi{3d}(m_{H\rmi{eff}}^\phi)+
I_\rmi{3d}(m_{U\rmi{eff}}^\phi)
\Bigr) \Bigl( 
-I(m_Q) + I_{n\neq 0} (0) 
\Bigr) \nn 
& & \hspace*{1cm} + 
I_\rmi{3d}(m_{H\rmi{eff}}^\phi)
I_\rmi{3d}(m_{U\rmi{eff}}^\phi)\Bigr] + {\cal O}(\delta_\phi m)^2,
\la{huq}
\ea
where $D,D_{n\neq 0}$ are from \eqs\nr{defD}, \nr{defDne0}
and we have used \eq\nr{Hlin},
we find that:
\bi
\item there are {\em linear terms} 
$\propto I_\rmi{3d}(m_{H\rmi{eff}}^\phi), I_\rmi{3d}(m_{U\rmi{eff}}^\phi)$
which however all get cancelled, 
when the choice in \eqs\nr{resum0}, \nr{resum2} is made for $\delta_r m_H^2, 
\delta_r m_U^2$.
\item there is an {\em infrared} sensitive contribution,
quadratic in the masses, from the Matsubara zero 
modes in the graphs $(HU)$, $(HUQ)$:
\ba
\left. (HU)+(HUQ)\right|_\rmi{IR} & = & 
6 (h_1^2-h_3^2 |\hat A|^2) 
I_\rmi{3d}(m_{H\rmi{eff}}^\phi)I_\rmi{3d}(m_{U\rmi{eff}}^\phi).
\la{HU}
\ea
The appearance of $h_1^2-h_3^2 |\hat A|^2$ corresponds to 
coupling constant resummation which we however do not discuss
in any detail here, since the corresponding effects are in principle
beyond the accuracy of \eq\nr{paramform}. Similarly, 
the graph $(HUQ)$ also produces terms 
of order $\sim h^4 |\hat A|^2 \phi^2$, 
again beyond \eq\nr{paramform}. 
\item finally, there are {\em ultraviolet} sensitive
(not from the zero modes) quadratic 
terms from the graphs $(HU)$, $(HQ)$:
\ba
\left. (HU)+(HQ)\right|_\rmi{UV} & = & 
-6 h_1^2 \delta_\phi m_{U\rmi{eff}}^2 
I_{n\neq 0}(m_{H}) D_{n\neq 0}(m_{U}) \nn
& & 
-3 h_2^2 \delta_\phi m_{Q_2}^2 I_{n\neq 0}(m_{H}) D(m_Q)
+ {\cal O}(\delta_\phi m)^4. \hspace*{2.0cm} \la{huuv}
\ea
\ei

To summarize, we have observed that all linear terms get
cancelled when the thermal counterterms are chosen according
to \eq\nr{Tct}. The remainder involves quadratic terms, 
which can either come from the ultraviolet  or the infrared.

\section{Next-to-leading order}
\la{2loop}

We next evaluate the 2-loop contributions from the remaining
graphs, and expand them again in $\delta_\phi m$; however,
these graphs do not involve contributions linear in $\delta_\phi m$. 
The graphs left are the sunsets $(HQQ), (HUU)$, as well 
as the 1-loop graphs $(H), (U), (Q)$, where the blobs
are now the bilinears obtained from the coupling 
constant counterterms after the shift of $H$~\footnote{Mass counterterms
do not contribute at the present order; terms proportional 
to $m_Q^2$ in them would, had we included self-interactions of the type 
$\sim (H^\dagger H)^2, (U^*_\alpha U_\alpha)^2$ in \eq\nr{lagr}.}. 

After an expansion in $\delta_\phi m$, we obtain
\ba
(H)+(U)+(Q) & = &  \frac{\phi^2}{2}\frac{1}{(4\pi)^2}\frac{1}{\epsilon}
\Bigl[9 (h_1^4+h_2^4) I_{n\neq 0}(m_H) + 6 h_1^4 I_{n\neq 0}(m_U) \nn
& &  + 
12 h_2^4 I(m_Q) 
\Bigr] + {\cal O}(\delta_\phi m)^3, \la{uuv} \\
(HQQ) & = &  - 3 h_2^4 \phi^2 H(m_Q,m_Q,0)  +
{\cal O} (\delta_\phi m)^3, \la{hqq} \\
(HUU) & = & - \fr32 h_1^4 \phi^2 
H(m_{H\rmi{eff}}^\phi,m_{U\rmi{eff}}^\phi,m_{U\rmi{eff}}^\phi). \la{HUU}
\ea
The numerical expression of $H(m_Q,m_Q,0)$ 
is discussed in appendix ~C. As to the graph $(HUU)$, 
on the other hand, we recall that it arises
completely from the zero Matsubara modes 
($H=H_\rmi{3d} + {\cal O}(m/T)$), and is thus 
a purely IR quantity~\cite{az}.

Adding all terms together from \eqs\nr{huuv}, \nr{uuv}, \nr{hqq}
and using \eqs\nr{Ine0}, \nr{defD}, \nr{defDne0}, \nr{Hfinite}, 
we obtain the 2-loop 
ultraviolet contribution to the 3d mass parameter, 
\ba
\delta_\rmi{2-loop}^\rmi{UV} m_{H\rmi{eff}}^2 \!\!\!\!\! & = & \!\!\!\!\!
h_2^4 \biggl[ 
-6 H_\rmi{vac} (m_Q,m_Q,0) + 12 \frac{1}{(4\pi)^2\epsilon}
I_\rmi{vac}(m_Q)
\biggr]
\nn
& + & \!\!\!\!\! \frac{T^2}{(4\pi)^2} \biggl\{
h_1^4 \biggl[ 
\fr34 \frac{1}{\epsilon} -\fr54 \ln\frac{\bmu^2}{\bmu_T^2} - 
3 \biggl( 
\ln\frac{3 T}{\bmu} + c \biggr) \biggr] \nn
& - & \!\!\!\!\! 
h_2^4 \biggl[ 
\fr34 \ln\frac{\bmu^2}{m_Q^2} 
+ 6{\cal I}_1  \Bigl( \frac{m_Q}{T}\Bigr)
\biggl(\ln\frac{\bmu^2}{m_Q^2}+2 \biggr) 
+\fr14 {\cal D} \Bigl( \frac{m_Q}{T}\Bigr)
+ 6 {\cal H}
\Bigl( \frac{m_Q}{T}\Bigr) \biggr]\biggr\}. \hspace*{0.5cm} \la{d2lmm}
\ea
Here the first line is a 2-loop vacuum renormalization 
correction of order $g^4 m_Q^2$, 
\be
\bmu_T = 4\pi e^{-\gamma_E} T \approx 7.0555T, \qquad  
c=\frac{1}{2}\biggl[\ln \frac{8\pi}{9}+\frac{\zeta'(2)}{\zeta(2)}-
2\gamma_E]\approx -0.34872274, \la{bmuT} 
\ee
and ${\cal I}_1$, ${\cal D}$, ${\cal H}$ are functions
defined in \eqs\nr{I1}, \nr{exp}, \nr{Hfin}. 

On the other hand, 
the IR sensitive part of the 
effective potential is, from \eqs\nr{HU}, \nr{HUU}, 
\ba
\delta_\rmi{2-loop}^\rmi{IR} V & = & 
6 (h_1^2-h_3^2|\hat A|^2) 
I_\rmi{3d}(m_{H\rmi{eff}}^\phi) I_\rmi{3d}(m_{U\rmi{eff}}^\phi) -
\fr32 h_1^4 \phi^2 
H_\rmi{3d}(m_{H\rmi{eff}}^\phi,m_{U\rmi{eff}}^\phi,m_{U\rmi{eff}}^\phi).
\hspace*{0.8cm}
\ea
The divergence in $H_\rmi{3d}$ (\eq\nr{H3}) cancels 
against that from \eq\nr{d2lmm}, $m_{H\rmi{eff}}^2\,\phi^2/2$.

Including also the 1-loop terms in \eq\nr{resum0}, 
we can now write down the 
complete mass parameter $m_{H\rmi{eff}}^2$ with accuracy 
$g^4m_Q^2, g^4 T^2$.
In order to do so, let us first note that 1-loop radiative
corrections generate couplings other than those in \eq\nr{lagr}, 
viz.\ 
\be
\delta {\cal L} =  h_4^2 H^\dagger H Q_\alpha^\dagger Q_\alpha + 
\lambda(H^\dagger H)^2 + ... ,
\ee
which we have to include in the discussion for a moment. 
The corresponding contribution in \eq\nr{resum0} is 
$\delta_r m_{H}^2 = 6 h_4^2 I(m_Q) + 6 \lambda I_{n\neq 0}(m_H^2)$.
Furthermore, in order to cancel spurious $\bmu$-dependences,
we should express the $\msbar$ parameters in terms of physical 
observables as in~\cite{generic}. In this paper we will not 
consider actual physical pole masses etc, but simply some 
finite physical scale independent parameters $()_\rmi{phys}$ 
which, dropping all terms beyond the accuracy of \eq\nr{paramform},  
we define through the following relations:
\ba
m_H^2(\bmu) & = & 
m_{H\rmi{phys}}^2 + \frac{3}{(4\pi)^2} \Bigl[
h_1^2 m_U^2  + 
(h_{2\rmi{phys}}^2 + h_3^2 |\hat A|^2) m_Q^2 \Bigr]
\biggl(\ln\frac{\bmu^2}{m_Q^2}+1 \biggr) \nn 
& & + h_2^4\Bigl[ 6 H_\rmi{vac}(m_Q,m_Q,0) - 
12 D_\rmi{vac}(m_Q) I_\rmi{vac}(m_Q)
\Bigr]_\rmi{finite part} , \\
h_1^2(\bmu) & = & 
h_{1\rmi{phys}}^2 + h_1^4 \frac{2}{(4\pi)^2} \ln\frac{\bmu^2}{m_Q^2}, \quad
h_2^2(\bmu) = 
h_{2\rmi{phys}}^2 + h_2^4 \frac{2}{(4\pi)^2} \ln\frac{\bmu^2}{m_Q^2}, \\
h_4^2(\bmu) & = & 
h_{4\rmi{phys}}^2 + h_2^4 \frac{1}{(4\pi)^2} \ln\frac{\bmu^2}{m_Q^2}, \quad
\lambda(\bmu) = 
\lambda_{\rmi{phys}} + \fr32 (h_1^4+h_2^4) 
\frac{1}{(4\pi)^2} \ln\frac{\bmu^2}{m_Q^2}, \hspace*{0.7cm}
\ea
where $H_\rmi{vac}(m_Q,m_Q,0)$, 
$ D_\rmi{vac}(m_Q)$, $ I_\rmi{vac}(m_Q)$,  
are defined in \eqs\nr{Hvacmm}, \nr{Dvacm}, \nr{Ivacm}. 
Moreover, let us now declare $h_{4\rmi{phys}}^2, \lambda_{\rmi{phys}}\sim 0$.
We then obtain the final 
expression for the 2-loop effective (bare)
mass parameter $m_{H\rmi{eff}}^2$
in the theory of \eq\nr{lagr}:
\ba
m_{H\rmi{eff}}^2 & = & m_{H\rmi{phys}}^2 -\frac{3}{(4\pi)^2}
h_1^2 m_U^2  \biggl( \ln\frac{m_Q^2}{\bmu_T^2} -1  \biggr) \nn
& + &  T^2 \biggl\{
\fr14 (h_{1\rmi{phys}}^2 - h_3^2 |\hat A|^2) + \fr32
(h_{2\rmi{phys}}^2 + h_3^2 |\hat A|^2)
{\cal I}_1 \Bigl( \frac{m_Q}{T} \Bigr) \nn 
& + & \frac{1}{(4\pi)^2} \biggl[ 
h_1^4 \biggl( 
\fr54 \ln\frac{\bmu_T^2}{m_Q^2} + \fr34\frac{1}{\epsilon} - 
3 \Bigl(\ln\frac{3 T}{\bmu} + c \Bigr) \biggr) \nn
& & -h_2^4 \biggl(
12 {\cal I}_1 \Bigl( \frac{m_Q}{T} \Bigr) + 
\fr14 {\cal D} \Bigl( \frac{m_Q}{T} \Bigr) + 
6 {\cal H} \Bigl( \frac{m_Q}{T} \Bigr)
\biggr)\biggr]\biggr\}.  \la{mHeff}
\ea

\section{High-$T$ and low-$T$ expansions}
\la{accuracy}

\begin{figure}[t]


\centerline{\epsfxsize=7cm
\epsfbox{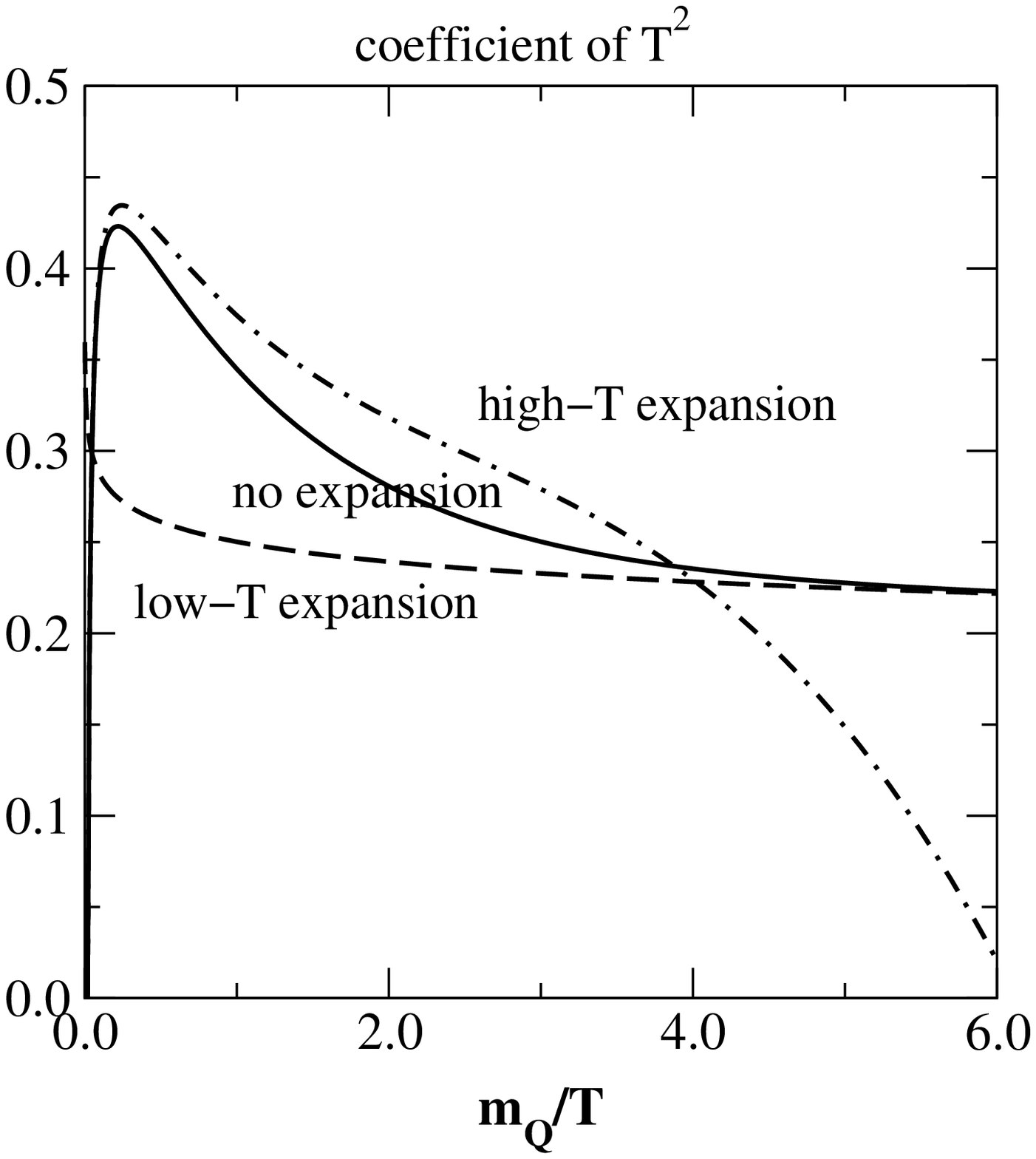}\hspace*{1cm}%
\epsfxsize=7cm\epsfbox{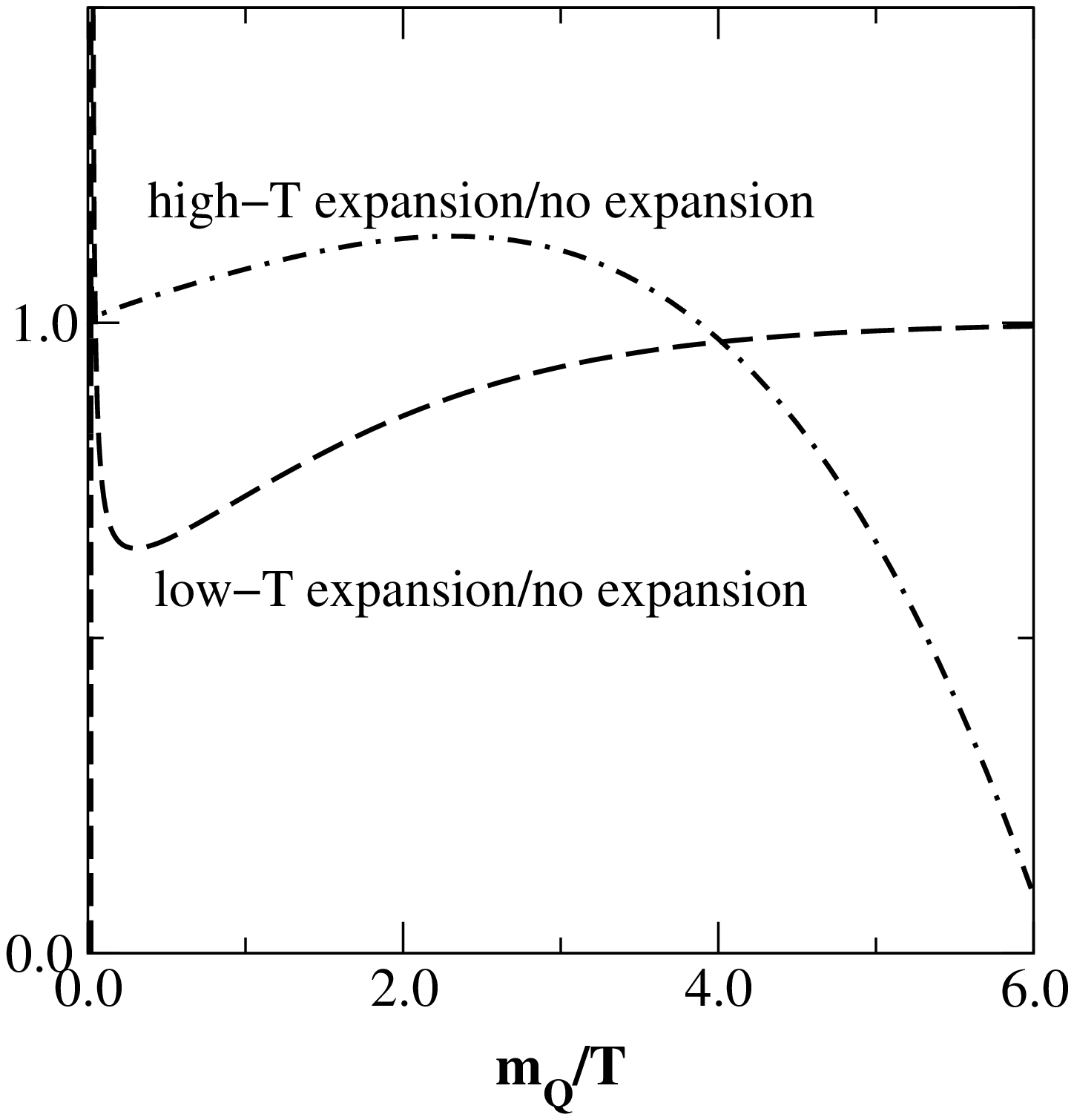}}


\caption[a]{Left: The coefficient of $T^2$ from \eq\nr{mHeff}, 
along with the high-$T$ and low-$T$ limits. The parameters 
have been chosen as explained in \se\ref{accuracy}. Right:
The ratio of the high-$T$ and low-$T$ limits to the
non-expanded result.}  
\la{fig:comp}
\end{figure}

We now wish to employ \eq\nr{mHeff} to estimate in a non-trivial
physical context the accuracy of the high and low temperature
expansions in $m_Q/T$. 
We can do this by inspecting the critical temperature $T_c$
of the phase transition. Let us recall that the leading 
(and next-to-leading in a gauge theory) terms in $T_c$ are 
perturbative~\cite{ar92}, thus ultraviolet dominated and
particularly sensitive to $m_Q/T$. Most of the physical 
characteristics of the phase transition, on the contrary, 
are infrared dominated and less sensitive to $m_Q/T$. 
We may also remind that in the MSSM the determination of $T_c$ 
is physically more 
important than in the Standard Model, since one has to address 
the question of whether other phase transitions could take place 
before the electroweak one, in particular a transition to the 
dangerous $U$-direction~\cite{bjls}--\cite{cms}.

The transition will take place when 
$m_{H\rmi{eff}}^2(\bmu=g^2T) = \# g^4T^2$, 
where $\#$ is some non-perturbative coefficient, to be determined
with lattice simulations. We shall keep the physical 
parameters $()_\rmi{phys}$ fixed and vary $m_Q/T$. 
It is then clear that the perturbative contribution 
to $T_c$ can equivalently be inspected by considering the 
finite part of the coefficient of $T^2$ in \eq\nr{mHeff}. 
We choose for simplicity
$h_{1\rmi{phys}},h_{2\rmi{phys}}\sim 1$, 
$h_3^2 |\hat A|^2 \sim  0$.  The high and low temperature limits
of ${\cal I}_1, {\cal D}, {\cal H}$ are given
in \eqs\nr{expI1}, \nr{expD}, \nr{Hfinlow}. 
To be in accordance with the limiting procedures
usually applied in the literature, we keep in the 
high temperature expansions terms up to logarithmic order, 
whereas in the low temperature expansions we simply 
replace the exponentially small corrections in 
\eqs\nr{expI1}, \nr{expD}, \nr{Hfinlow} with zero. 

The numerically evaluated full expression for the 
coefficient of $T^2$, as well as a comparison of 
the high and low temperature expanded versions thereof with
the full result,  are plotted in \fig\ref{fig:comp}.
We observe that the high temperature expansion gives 
typically too large a coefficient of the 
$T^2$-term, leading to too small a $T_c$~\footnote{Numerically 
the relative effect is larger here than in the realistic MSSM,  
since in that case there are other terms in the coefficient
of $T^2$ (such as gauge bosons) for which the high temperature
expansion should work perfectly.}. 
With the low-temperature 
expansion, on the other hand, $T_c$ is slightly too large.   
Furthermore, we observe that while naively one might have expected 
the crossover between the high and low temperature 
regimes to be close to the first non-zero Matsubara 
frequency at $m_Q\sim 2\pi T$, the low
temperature expansion is in fact 
perfectly sufficient already at $m_Q \gsim 3T$, which
is the case for realistic values $m_Q\gsim 300$ GeV.
The fact that the high temperature expansion converges
relatively poorly as early as at $m_Q/T\sim 2$ is due 
particularly to the 2-loop function
${\cal H}$, whose behaviour is shown in \fig\ref{fig:hnum}.

\section{Conclusions}
\la{concl}

In this paper, we have pointed out that standard thermal 
resummations should be extended in two different ways, when
one goes from the Standard Model to a general MSSM. First of all, 
the left-handed stop $m_Q$ is typically of the order of magnitude
$\sim 2\pi T$. Then it cannot, a priori, be treated either in 
the high or in the low temperature expansion, but a more general
function appears. Second, the presence of dimensionful trilinear
couplings leads to the emergence of new ``linear'' terms 
coming from the scalar sunset diagrams. The results for the 
scalar thermal counterterms including both of these effects in the 
model of \eq\nr{lagr} are shown in \eqs\nr{resum0}, \nr{resum2}, while the
scalar contributions to the Debye masses 
are shown in \eqs\nr{debye1}, \nr{debye}. In an 
effective theory approach such as the one followed in~\cite{cp}
(in~\cite{cp} it was assumed that $m_Q \gg 2\pi T$, but the 
procedure can be extended to $m_Q \sim 2\pi T$ in a 
straightforward way),
all these effects of course arise automatically, whereas in a 
direct computation of the 2-loop effective potential they
should be explicitly taken into account. 

In the framework of effective field theories, we have also 
extended the resummation for the Higgs mass parameter
to the next order beyond
the effects described above. The mass parameter thus determined, 
including corrections of order $\sim g^4T^2$, could be used for
a precise estimation of the critical temperature
of the corresponding
electroweak phase transition using 3d lattice
Monte Carlo simulations. Let us stress that the only change with
respect to previous effective 3d theories is in the expressions
for the effective parameters, not in the functional form of
the theory.  

Using these results, we have estimated the accuracy of 
the high and low temperature expansions used previously in the
literature. Inspection of the critical temperature suggests that 
the low temperature expansion, whereby all finite temperature
contributions from heavy particles are simply left out, works
well already at $m \gsim 3T$ for bosonic particles. Thus it 
should be completely clear that for the values $m_Q\sim 1$ TeV
of interest for obtaining a strong phase transition with 
experimentally allowed Higgs masses in the MSSM, 
the $Q$-field can simply
be left out in all finite temperature contributions. 

The present results could clearly be extended in many directions. 
First of all, the restricted model we have employed here can 
be extended to the full MSSM with gauge fields and fermions
in a straightforward way. 
Second, we have shown that the evaluation of the integrals 
appearing in the perturbative 2-loop effective potential is 
numerically feasible without any further temperature expansions --- 
thus the complete 2-loop potential of the MSSM could in principle be 
computed, extending thus the results 
of~\cite{bjls,cqw}, \cite{cm}--\cite{cms}, \cite{e}. 
Third, we have here considered explicitly 
only the effects of a heavy $m_Q$, while in the MSSM many other
mass parameters could be heavy as well.
In particular, $M_2,\mu$ related to the gaugino and Higgsino mass matrices 
and also relevant for providing sources of CP violation can have values
for which neither high nor low temperature expansions are applicable. 
In the effective theory approach $M_2,\mu$ can be easily included 
at 1-loop level without any temperature expansions~\cite{cp}, but
this could now be extended to the 2-loop level. The accuracy
of previous approximations with respect to the contributions from 
the second Higgs doublet  with $m_A\gsim 100$ GeV
could also be explicitly checked. 
Finally, we assumed that the trilinear couplings are not exceedingly 
large, $|\hat A|^2\lsim g^2$, an assumption 
which could be relaxed. 

We believe that as long as the existence of a Higgs particle with 
MSSM type couplings lighter than about 110 GeV is not experimentally 
excluded, these are worthwhile questions to consider. 

\section*{Acknowledgements}

The work of  M.~Laine was partly supported by the TMR network {\em Finite
Temperature Phase Transitions in Particle Physics}, EU contract no.\
FMRX-CT97-0122. M.~Losada thanks A.~Nieto for useful discussions during
early stages of this work.


\appendix
\renewcommand{\theequation}{\Alph{section}.\arabic{equation}}

\section{The tadpole}

For completeness, let us review here some properties of the tadpole integral,  
\be
I(m) = T\sum_n \int\! \frac{d^{3-2\epsilon}p}{(2\pi)^{3-2\epsilon}}
\frac{1}{p_0^2 + p^2 + m^2},  \la{tadpole}
\ee
where $p_0 = p_b \equiv 2\pi nT$ for bosons, 
$p_0 = p_f \equiv \pi T(2n+1)$ for fermions. 
We will need two types of subdivisions of $I(m)$. In the first 
case, relevant for all fermions and heavy bosons,
we write $I(m) = I_\rmi{vac}(m) + I_T(m)$, where $I_T$ vanishes at $T=0$. 
In the second case, relevant for light bosons 
($m^2\sim (g T)^2$), we separate the contribution from the 
Matsubara zero mode into $I_\rmi{3d}(m)$, writing 
$I(m) = I_\rmi{3d}(m) + I_{n\neq 0}(m)$. We denote
\ba
& & n_b(\omega) = \frac{1}{e^{\beta \omega} - 1}, \quad
n_f(\omega) = \frac{1}{e^{\beta \omega} + 1}, \\ 
& & 
\omega_{p,i} = (p^2 + m_i^2 ) ^{1/2}, \quad
\hat \omega_{p} = (p^2 + (m/T)^2 ) ^{1/2}, \\
& & 
I_{T{,b(f)}} (m_i) = \int \! \frac{d^{3-2\epsilon} p }{(2\pi)^{3-2\epsilon}}
\frac{n_{b(f)}
(\omega_{p,i})}{\omega_{p,i}}.
\ea
When the superscript $()_{b,f}$
is left out from $I$, we assume the bosonic case. 

\paragraph{A heavy mass in the loop.}
Writing $I_b(m) = I_\rmi{vac}(m) + I_T(m)$, we get
\ba
I_\rmi{vac}(m) & = & -\mu^{-2\epsilon}\frac{m^2}{(4\pi)^2}\biggl(
\frac{1}{\epsilon} + \ln\frac{\bmu^2}{m^2}+1
\biggr), \la{Ivacm} \\ 
I_T(m) & = & \fr12 \mu^{-2\epsilon}T^2 \biggl\{
\Bigl[1 + \epsilon\Bigl(2- 2\ln 2 + \ln\frac{\bmu^2}{T^2 }\Bigr) \Bigr]
{\cal I}_1 \Bigl(\frac{m}{T}\Bigr) 
- \epsilon {\cal I}_2 \Bigl(\frac{m}{T}\Bigr)\biggr\}, \la{IT} \\
{\cal I}_1 \Bigl(\frac{m}{T}\Bigr) & = &  \frac{1}{\pi^2} 
\int_0^\infty \! dp\, p^2 \frac{n_b(\hat \omega_p)}{\hat \omega_p},
\la{I1} \\
{\cal I}_2 \Bigl(\frac{m}{T}\Bigr) & = &  \frac{1}{\pi^2} 
\int_0^\infty \! dp\, p^2 \ln p^2\frac{n_b(\hat \omega_p)}{\hat \omega_p}.
\la{I2}
\ea
The limiting values are
\ba
{\cal I}_1 (y) & \stackrel{y\ll 1}{=} & \frac{1}{6}-\frac{y}{2\pi}+
\frac{y^2}{8\pi^2}\Bigl(1+2\ln\frac{4\pi}{y}-2\gamma_E \Bigr),  \qquad
\stackrel{y\gg 1}{=}  
\sqrt{\frac{y}{2\pi}} \frac{e^{-y}}{\pi}, \la{expI1} \\
{\cal I}_2 (y) & \stackrel{y\ll 1}{=} & \frac{1}{3} 
\Bigl[ 1-\gamma_E +\frac{\zeta'(2)}{\zeta(2)}\Bigr]-
\frac{y}{\pi}\ln y, \quad 
\stackrel{y\gg 1}{=} 
\sqrt{\frac{y}{2\pi}} \frac{e^{-y}}{\pi} 
\Bigl( \ln y + 2 -\gamma_E -\ln 2\Bigr), \hspace*{1.3cm}
\ea
where $\gamma_E = 0.57721566, \zeta'(2)/\zeta(2)= -0.56996099$.

\paragraph{A light mass in the loop.}
Writing $I(m)=I_{n\neq 0}(m) + I_\rmi{3d}(m)$, we obtain
\ba
I_{n\neq 0}(m) & = & 
\mu^{-2\epsilon}\frac{T^2}{12}(1+\epsilon \imath_\epsilon) - 
\mu^{-2\epsilon}\frac{m^2}{(4\pi)^2}\biggl(\frac{1}{\epsilon} + 
\ln\frac{\bmu^2}{\bmu_T^2}\biggr) + 
{\cal O}(\epsilon^2, \epsilon m^2, m^4), \la{Ine0} \\ 
I_\rmi{3d}(m) & = & -\mu^{-2\epsilon}  \frac{mT}{4\pi} 
\biggl[ 1+ \epsilon\Bigl(
\ln\frac{\bmu^2}{m^2} + 2 - 2\ln 2
\Bigr) \biggr]
+ {\cal O}(\epsilon^2). \la{I3d}
\ea
Here $\bmu_T$ is from \eq\nr{bmuT} and~\cite{ae}
\ba
\imath_\epsilon & = &\ln\frac{\bmu^2}{T^2}+2 \gamma_E-2\ln 2-
2\frac{\zeta'(2)}{\zeta(2)}. \la{ie}
\ea

\paragraph{The fermionic tadpole.}
Using the standard trick, the fermionic tadpole can be 
expressed in terms of the bosonic tadpole:
\ba
I_f(m) &  = &  I_\rmi{vac}(m) + I_{T,f}(m),  \\ 
I_{T,f}(m) & = & 2 I_{T/2,b}(m)-I_{T,b}(m) 
= 2^{-1+2\epsilon} I_{T,b}(2m)-I_{T,b}(m).
\ea
Defining ${\cal I}_{1,f}, {\cal I}_{2,f}$ as in \eqs\nr{I1}, \nr{I2}
but with $n_f$ instead of $n_b$, we obtain
\be
I_{T,f}(m) = \fr12 \mu^{-2\epsilon} T^2 \biggl\{
- \Bigl[1 + \epsilon\Bigl(2- 2\ln 2 + \ln\frac{\bmu^2}{T^2 }\Bigr) \Bigr]
{\cal I}_{1,f} \Bigl(\frac{m}{T}\Bigr) 
+ \epsilon {\cal I}_{2,f} \Bigl(\frac{m}{T}\Bigr)\biggr\}.
\ee
The high and low temperature expansions of ${\cal I}_{1,f}, {\cal I}_{2,f}$
can be obtained from those of ${\cal I}_{1}, {\cal I}_{2}$ by noting that
\ba
{\cal I}_{1,f}\Bigl(\frac{m}{T}\Bigr) & = & 
{\cal I}_{1}\Bigl(\frac{m}{T}\Bigr)-\fr12 
{\cal I}_{1}\Bigl(\frac{2 m}{T}\Bigr), \\ 
{\cal I}_{2,f}\Bigl(\frac{m}{T}\Bigr) & = & 
{\cal I}_{2}\Bigl(\frac{m}{T}\Bigr)-\fr12
{\cal I}_{2}\Bigl(\frac{2m}{T}\Bigr)+\ln 2\;
{\cal I}_{1}\Bigl(\frac{2m}{T}\Bigr).
\ea

\section{The bubble}

Let us then consider the 1-loop ``bubble'' diagram with two propagators, 
\be
D_{{b(f)}}(m_1,m_2)  =  \Tint{P_{b(f)}}
\frac{1}{P^2+m_1^2}\frac{1}{P^2+m_2^2}  =  
\frac{1}{m_1^2-m_2^2} \Bigl[ I_{b(f)}(m_2) - 
I_{b(f)}(m_1)\Bigr], \la{Dbf}
\ee
where $P_{b(f)}=(p_{b(f)},\vec{p})$ and
we have taken the external momentum to zero. We can again write
\ba
D_{b(f)}(m_1,m_2) & = &  D_\rmi{vac}(m_1,m_2) + D_{T{,b(f)}}(m_1,m_2), \\
D_\rmi{vac}(m_1,m_2) & = &  
\frac{\mu^{-2\epsilon}}{(4\pi)^2}\biggl(
\frac{1}{\epsilon} + \ln\frac{\bmu^2}{m_1 m_2}+1 -
\frac{m_1^2+m_2^2}{m_1^2-m_2^2} \ln \frac{m_1}{m_2}
\biggr), \\
D_{T{,b(f)}}(m_1,m_2) & = &  \pmm\mu^{-2\epsilon}
\frac{T^2}{2}\frac{1}{m_1^2-m_2^2}
\Bigl[ {\cal I}_{1,b(f)}\Bigl(\frac{m_2}{T}\Bigr) -
{\cal I}_{1,b(f)}\Bigl(\frac{m_1}{T}\Bigr)\Bigr] +{\cal O}(\epsilon).
\ea
The special case $m_1=m_2$    
gives the derivative of $I(m)$ 
with respect to $m^2$:
\ba
D(m) & \equiv & -\frac{d I(m)}{d m^2} = D_\rmi{vac}(m) + D_T(m), 
\la{defD} \\
D_\rmi{vac}(m) & = & 
\frac{\mu^{-2\epsilon}}{(4\pi)^2}
\biggl(\frac{1}{\epsilon}  +\ln\frac{\bmu^2}{ m^2} \biggr), \la{Dvacm} \\
D_T(m) & = & \frac{\mu^{-2\epsilon}}{(4\pi)^2} 
{\cal D}\Bigl(\frac{m}{T} \Bigr) + 
{\cal O}(\epsilon), \\
{\cal D}\Bigl(\frac{m}{T} \Bigr) & = & 4 
\int_0^\infty \! dp \frac{n_b(\hat \omega_p)}{\hat\omega_p}, \la{exp}
\ea
with
\be
{\cal D} (y) \stackrel{y\ll 1}{=} \frac{2\pi}{y}+
2\ln\frac{y}{4\pi}+2\gamma_E, \qquad
\stackrel{y\gg 1}{=} 
2 \sqrt{\frac{2\pi}{y}} {e^{-y}}. \la{expD}
\ee
We also need the derivative of $I_{n\neq 0}(m)$ 
with respect to $m^2$: 
\be
D_{n\neq 0}(m) \equiv -\frac{d I_{n\neq 0}(m)}{d m^2}  =
\frac{\mu^{-2\epsilon}}{(4\pi)^2}
\biggl(\frac{1}{\epsilon}  +\ln\frac{\bmu^2}{\bmu_T^2} \biggr)+ 
{\cal O}(\epsilon,m^2). \la{defDne0}
\ee

\section{The sunset}

Let us then
consider the bosonic and fermionic 2-loop sunset diagrams
\ba
H_{b(f)}(m_1,m_2,m_3) & = & \Tint{P_{b(f)}}\Tint{Q_{b(f)}} 
\frac{1}{P^2+m_1^2}\frac{1}{Q^2+m_2^2}\frac{1}{(P+Q)^2+m_3^2}.
\la{sunset}
\ea
In the limit $m_i/T \ll 1$, it is known that~\cite{ae,az} 
\ba
H_b(m_1,m_2,m_3) & = &  \mu^{-4\epsilon} \frac{T^2}{(4\pi)^2} 
\biggl(\frac{1}{4\epsilon} +\ln\frac{\bmu}{m_1+m_2+m_3} + \fr12 \biggr)
+ {\cal O} (m_i T), \la{H3} \\
H_f(m_1,m_2,m_3) & = &  {\cal O} (m_i T). 
\ea
Our objective here is to compute these 
diagrams in the case of general $m_i$. 
We are aware of previous results in this direction 
in~\cite{Parwani:1992gq,Jakovac:1996ue}.

\paragraph{General case.}
The method we employ for evaluating $H_b,H_f$
follows the standard procedure
(see, e.g.,\ \cite{Parwani:1992gq,Jakovac:1996ue,Bugrii:1995vn}). 
The twofold sum over the Matsubara
modes is first written as a threefold sum with a Kronecker delta, 
and the delta is then written as 
$\delta(p_0) = T \int_0^\beta dx \exp(i p_0 x)$.
The sums can now be performed, 
\be
T\sum_{p_{b(f)}} \frac{e^{i p_{b(f)} x}}{p_{b(f)}^2 + \omega_i^2} = 
\frac{n_{b(f)}(\omega_i)}{2\omega_i} \Bigl[ 
e^{(\beta-x)\omega_i} \pmm e^{x \omega_i}
\Bigr].
\ee
The integral over $x$ is then very simple. The
outcome can be organized in a transparent form, 
when different types of contributions are identified 
with known expressions;
the same result could also have been obtained from the rules 
of the real time formalism, as noted for the 3-loop bosonic basketball 
diagram in~\cite{Andersen:2000zn}.
In the remaining integral over the spatial vectors $\vec{p}, \vec{q}$, 
we can perform at least the integration over 
$z=\vec{p}\cdot\vec{q}/(|\vec{p}||\vec{q}|)$, leaving for numerics 
at most a rapidly convergent 2d integral over $p\equiv |\vec{p}|$,
$q\equiv |\vec{q}|$. 

Let us denote 
\ba 
\Pi(Q^2;m_1^2,m_2^2) & = & 
\int \! \frac{d^{4-2\epsilon} P }{(2\pi)^{4-2\epsilon}}
\frac{1}{[P^2+m_1^2][(P+Q)^2 + m_2^2]}, \\
f_{p,q}(m_1,m_2;m_3) & = & \ln \left|
\frac{4(p^2+m_1^2)(q^2+m_2^2)-(m_1^2+m_2^2-m_3^2-2p q)^2}
{4(p^2+m_1^2)(q^2+m_2^2)-(m_1^2+m_2^2-m_3^2+2p q)^2}\right|. \la{fpq}
\ea
The explicit expression for $\Pi(Q^2;m_1^2,m_2^2)$, 
often denoted by $B_0$, is well known:
\ba
\Pi(Q^2;m_1^2,m_2^2) \!\!\!\! & = &  \!\!\!\! \frac{\mu^{-2\epsilon}}{(4\pi)^2}
\biggl[
\frac{1}{\epsilon} + \ln\frac{\bmu^2}{m_1 m_2}+1 -
\frac{m_1^2+m_2^2}{m_1^2-m_2^2} \ln \frac{m_1}{m_2} + 
F_E(Q^2; m_1^2,m_2^2)
\biggr], \mbox{\hspace*{0.8cm}} \la{FE1} \\
F_E(Q^2; m_1^2,m_2^2) \!\!\!\! & = & \!\!\!\!  
1+\frac{m_1^2+m_2^2}{m_1^2-m_2^2} \ln \frac{m_1}{m_2}+ 
\frac{m_1^2-m_2^2}{Q^2} \ln \frac{m_1}{m_2} \nn
\!\!\!\! & + & \!\!\!\!\!  
\frac{1}{Q^2} 
\sqrt{(m_1+m_2)^2\! +\! Q^2}
\sqrt{(m_1-m_2)^2\! +\! Q^2} \ln
\frac{1-\sqrt{\frac{(m_1-m_2)^2+Q^2}{(m_1+m_2)^2+Q^2}}}
{1+\sqrt{\frac{(m_1-m_2)^2+Q^2}{(m_1+m_2)^2+Q^2}}}.
\la{FE}
\ea
The absolute value inside the logarithm in $f_{p,q}$ 
in \eq\nr{fpq} means that 
we take the real part of the expression; the imaginary part would 
anyway cancel against $\im\Pi$.  

With this notation, we obtain
\ba
& & \!\!\!\!\!\!\!\!\!\!\!\! H_b(m_1,m_2,m_3) =  
H_\rmi{vac}(m_1,m_2,m_3) \nn 
& & 
+ \sum_{i\neq j \neq k}
I_{T,b}(m_i)\re\Pi(-m_i^2;m_j^2,m_k^2) \nn
& & +  \sum_{i\neq j \neq k} \frac{\mu^{-4\epsilon}}{32\pi^4} 
\int_0^\infty \! dp\, p \int_0^\infty \!dq\, q
\frac{n_b(\omega_{p,i})}{\omega_{p,i}} 
\frac{n_b(\omega_{q,j})}{\omega_{q,j}} 
f_{p,q}(m_i,m_j;m_k), \la{Hb} \\
& & \nn
& &  \!\!\!\!\!\!\!\!\!\!\!\! H_f(m_1,m_2,m_3) =  
H_\rmi{vac}(m_1,m_2,m_3) \nn
& & + I_{T,b}(m_3)\re\Pi(-m_3^2;m_1^2,m_2^2) 
- \sum_{i\neq j} I_{T,f}(m_i)\re\Pi(-m_i^2;m_j^2,m_3^2) \nn
& & + \frac{\mu^{-4\epsilon}}{32\pi^4} 
\int_0^\infty \! dp\, p 
\int_0^\infty \! dq\, q
\frac{n_f(\omega_{p,1})}{\omega_{p,1}} 
\frac{n_f(\omega_{q,2})}{\omega_{q,2}} 
f_{p,q}(m_1,m_2;m_3) \nn 
& & - \sum_{i\neq j} 
\frac{\mu^{-4\epsilon}}{32\pi^4} 
\int_0^\infty \! dp\, p
\int_0^\infty \! dq\, q
\frac{n_b(\omega_{p,3})}{\omega_{p,3}} 
\frac{n_f(\omega_{q,i})}{\omega_{q,i}} 
f_{p,q}(m_3,m_i;m_j), \la{Hf}
\ea
where 
$\sum_{i\neq j\neq k}\equiv \sum_{(i,j,k)=(1,2,3),(2,3,1),(3,1,2)}$,
$\sum_{i\neq j}\equiv \sum_{(i,j)=(1,2),(2,1)}$, 
and the zero temperature contribution is 
\be
H_\rmi{vac}(m_1,m_2,m_3) 
= \int \! \frac{d^{4-2\epsilon}P}{(2\pi)^{4-2\epsilon}}
\int \! \frac{d^{4-2\epsilon}Q}{(2\pi)^{4-2\epsilon}}
\frac{1}{P^2+m_1^2}\frac{1}{Q^2+m_2^2}\frac{1}{(P+Q)^2+m_3^2}. 
\la{zeroT}
\ee
The only ultraviolet divergences are in 
$H_\rmi{vac}(m_1,m_2,m_3)$,
which has $1/\epsilon^2,1/\epsilon$ poles,  
and in the $\Pi$'s, which have the pole 
$\mu^{-2\epsilon}/(16\pi^2\epsilon)$, cf.\ \eq\nr{FE1}.

Let us also mention a few words about the numerical evaluation 
of the 2d integrals involving $f_{p,q}(m_1,m_2;m_3)$, left to be carried 
out in \eqs\nr{Hb}, \nr{Hf}. These integrals are of course
well-defined and finite. However, if $m_3 < |m_1-m_2|$ or 
$m_3 > m_1+m_2$, they involve integration over logarithmic 
singularities. In our numerics, we found that the integration 
is more effective if we factorise out the singularities explicitly.  
Suppose we, for instance, first perform the integral over 
$q$. If $m_1\neq 0$, we write
\ba
& & a = \frac{p}{2m_1^2}(m_3^2-m_1^2-m_2^2), \\
& & b = \frac{1}{2m_1^2}(p^2+m_1^2)^{1/2}\Bigl(
m_3^4-2m_3^2(m_1^2+m_2^2)+(m_1^2-m_2^2)^2
\Bigr)^{1/2}, \\
& & 
f_{p,q}(m_1,m_2;m_3)= 
\ln\left|\frac{(q-a-b)(q+b-a)}{(q+a+b)(q+a-b)}\right|, 
\ea
and there are then
singularities at $q=|a+b|,|a-b|$. If $m_1=0$, we write
\be 
q_0 = \frac{4m_2^2p^2-(m_3^2-m_2^2)^2}{4p(m_3^2-m_2^2)}, \quad
f_{p,q}(0,m_2;m_3)=\ln\left|\frac{q-q_0}{q+q_0}\right|,  
\ee
and there is a singularity at $q=|q_0|$. If $m_2=m_3$, $f_{p,q}(0,m;m)=0$.

\paragraph{One heavy, one light mass.}
Let us now consider in more detail some special cases of 
$H_b(m_1,m_2,m_3)$ needed in the main part of this paper. 
For the consideration in \se\ref{linear},
we need to know how $H_b(m_{H},m_{U},m_Q)$
behaves for small $m_{H},m_{U}$. We claim that
there is a linear term $\propto m_{H},m_{U}$ (modulo logarithms). 
Since the result is symmetric in $m_{H},m_{U}$, 
it is enough to consider $m_{H}\ll m_Q$.

\begin{figure}[t]


\centerline{\epsfxsize=7cm
\epsfbox{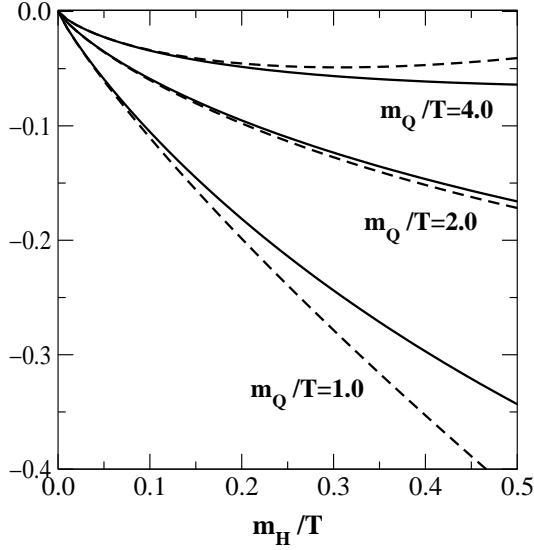}}


\caption[a]{The finite part of  $H_b(m_H,m_Q,0)-H_b(0,m_Q,0)$ 
(the zero temperature contribution has been subtracted) with solid lines, 
compared with the ``linear term'' in \eq\nr{Hlin} with dashed lines.  
The results have been divided by $T^2/(4\pi)^2$.}  
\la{fig:hmM0}
\end{figure}

Non-analytic terms can only arise from a zero Matsubara mode. Thus, 
\be
H_b \stackrel{m_H\ll m_Q}{\sim}
\int\! \frac{d^{3-2\epsilon}p}{(2\pi)^{3-2\epsilon}} \frac{1}{p^2+m_H^2}
\Tint{Q} \frac{1}{q_0^2 + (p+q)^2 + m_U^2} \frac{1}{q_0^2+q^2+m_Q^2}.
\ee
Let us denote the latter integral by $\Pi(p)$. It is then
obvious that the leading behaviour must be
\ba
H_b & \stackrel{m_H\ll m_Q}{\sim} & 
\int\! \frac{d^{3-2\epsilon}p}{(2\pi)^{3-2\epsilon}} \frac{1}{p^2+m_H^2} 
\Pi(0) = I_\rmi{3d}(m_H) \frac{1}{m_Q^2-m_U^2}
\Bigl[I(m_U)-I(m_Q)\Bigr] \nn
& = & \frac{1}{m_Q^2} I_\rmi{3d}(m_H) 
\Bigl[-I(m_Q) + I_\rmi{3d}(m_U) + I_{n\neq 0}(0) \Bigr]
\biggl(1+{\cal O}\Bigl(\frac{m_U^2}{m_Q^2},\frac{m_U^2}{T^2} \Bigr) \biggr).
\hspace*{1cm}
\la{Hlin}
\ea
Indeed, the remainder, 
\be
\sim \int\! \frac{d^{3-2\epsilon}p}{(2\pi)^{3-2\epsilon}} \frac{1}{p^2+m_H^2} 
\Bigl[\Pi(p)-\Pi(0)\Bigr],
\ee
behaves at small $p$ as $\int\! dp\, p^4/(p^2+m_H^2)$. This is IR finite
even after expanding in $m_H^2$, therefore there cannot be any further 
linear contributions. 

In order to verify this behaviour explicitly, 
we set $m_U=0$ and compare the finite
part of \eq\nr{Hlin} with a numerical evaluation of the finite
part of $H_b(m_H,m_Q,0)$ in \eq\nr{Hb}.
We fix $\bmu=T$, and choose $m_Q/T=1.0,2.0,4.0$. The result is shown 
in \fig\ref{fig:hmM0}. Note that in \eq\nr{Hlin}, one must 
include a contribution $\sim m_H\ln m_H$, arising when the 
${\cal O}(\epsilon)$ part of $I_\rmi{3d}(m_H)$ (cf.\ \eq\nr{I3d})
combines with the $1/\epsilon$ pole in $I(m_Q)$. {}From 
the perfect agreement at small $m_H/T$ in \fig\ref{fig:hmM0}, 
we conclude that for $m_H/T \ll 1$ the
behaviour is indeed according to \eq\nr{Hlin}.

\paragraph{Two equal heavy masses.}
Finally, let us consider the special case 
needed in \se\ref{2loop}, $H_b(m,m,0)$
(cf.\ \eq\nr{hqq}). The $T=0$ part in \eq\nr{zeroT}, 
related to 2-loop vacuum renormalization of $m_H^2(\bmu)$, is
\be
H_\rmi{vac}(m,m,0) = 
-\mu^{-4\epsilon} \frac{m^2}{(4\pi)^4} 
\left( \frac{\bmu^2}{m^2} \right)^{2\epsilon} 
\biggl( \frac{1}{\epsilon^2} + \frac{3}{\epsilon} + 7 + \frac{\pi^2}{6} + 
{\cal O}(\epsilon)
\biggr).
\la{Hvacmm} 
\ee
Using \eq\nr{IT} as well as the simple expressions for 
$\Pi(0;m^2,m^2), \Pi(-m^2;m^2,0)$ obtained from 
\eqs\nr{FE1}, \nr{FE}, 
we then get from \eq\nr{Hb}
\ba
& & \!\!\!\!\!\!\!\!\!\!\!\! H_b(m,m,0) = 
H_\rmi{vac}(m,m,0) 
+ \mu^{-4\epsilon} \frac{T^2}{(4\pi)^2} \biggl[
\biggl(\frac{1}{12} + {\cal I}_1\Bigl(\frac{m}{T}\Bigr) \biggr)
\biggl(\frac{1}{\epsilon} + \ln\frac{\bmu^2}{T^2} + \ln\frac{\bmu^2}{m^2}
\biggr) \nn 
& & + (4-2\ln2){\cal I}_1\Bigl(\frac{m}{T}\Bigr) 
-{\cal I}_2\Bigl(\frac{m}{T}\Bigr)
+ \fr16 \biggl(\gamma_E - \ln 2 - \frac{\zeta'(2)}{\zeta(2)}\biggr)  
+ {\cal H}\Bigl(\frac{m}{T}\Bigr)
\biggr],  \la{Hfinite}
\ea
where ${\cal I}_1, {\cal I}_2$ are from \eq\nr{I2}, and 
\be
{\cal H}\Bigl(\frac{m}{T}\Bigr) = \frac{2}{\pi^2}
\int_0^\infty \! dp\, p
\int_0^p \! dq\, q
\frac{n_b(\hat \omega_p)}{\hat \omega_p} 
\frac{n_b(\hat \omega_q)}{\hat \omega_q} \ln\frac{p+q}{p-q}.
\la{Hfin}
\ee

\begin{figure}[t]


\centerline{\epsfxsize=7cm
\epsfbox{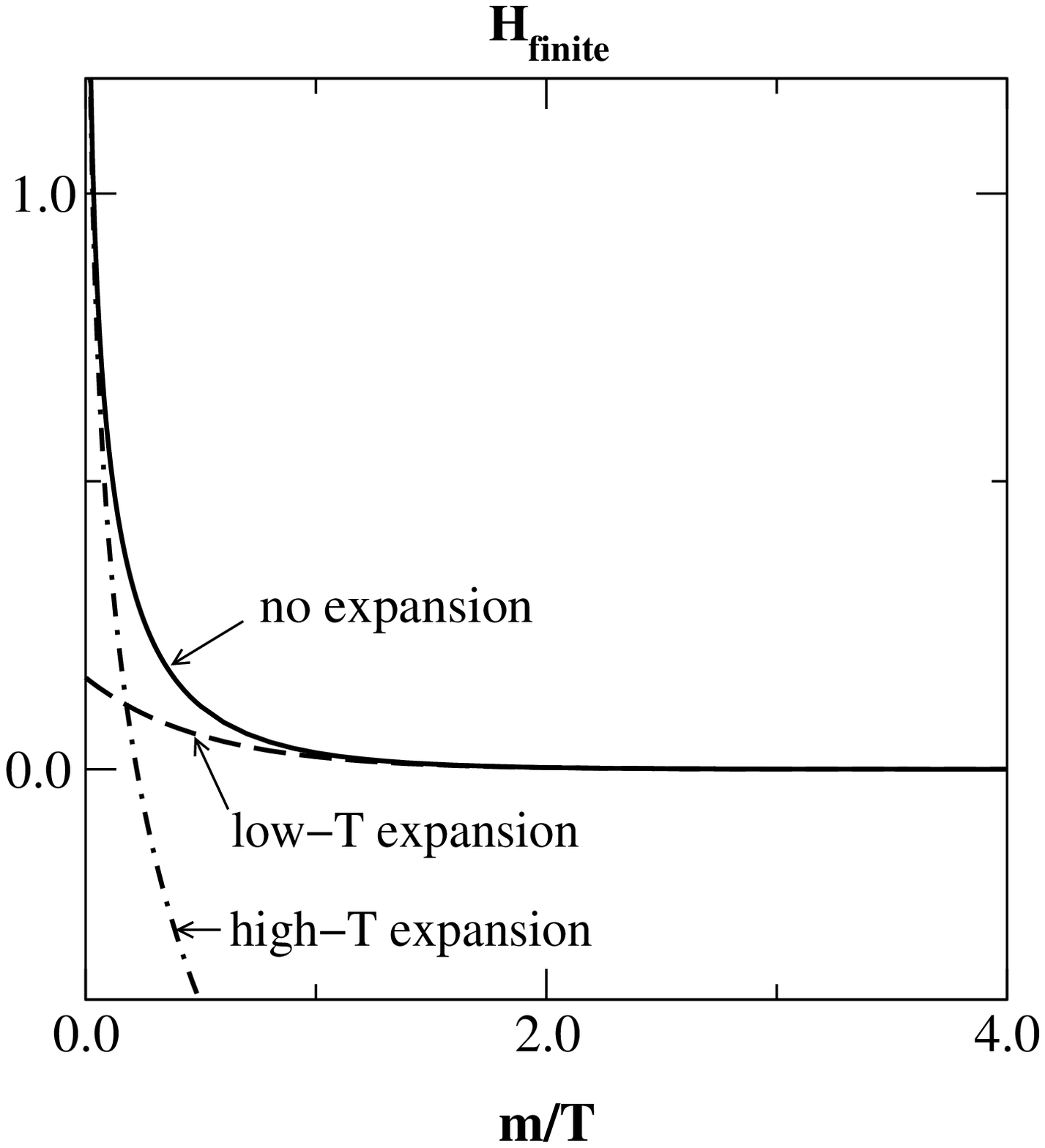}\hspace*{1cm}
\epsfxsize=7cm%
\epsfbox{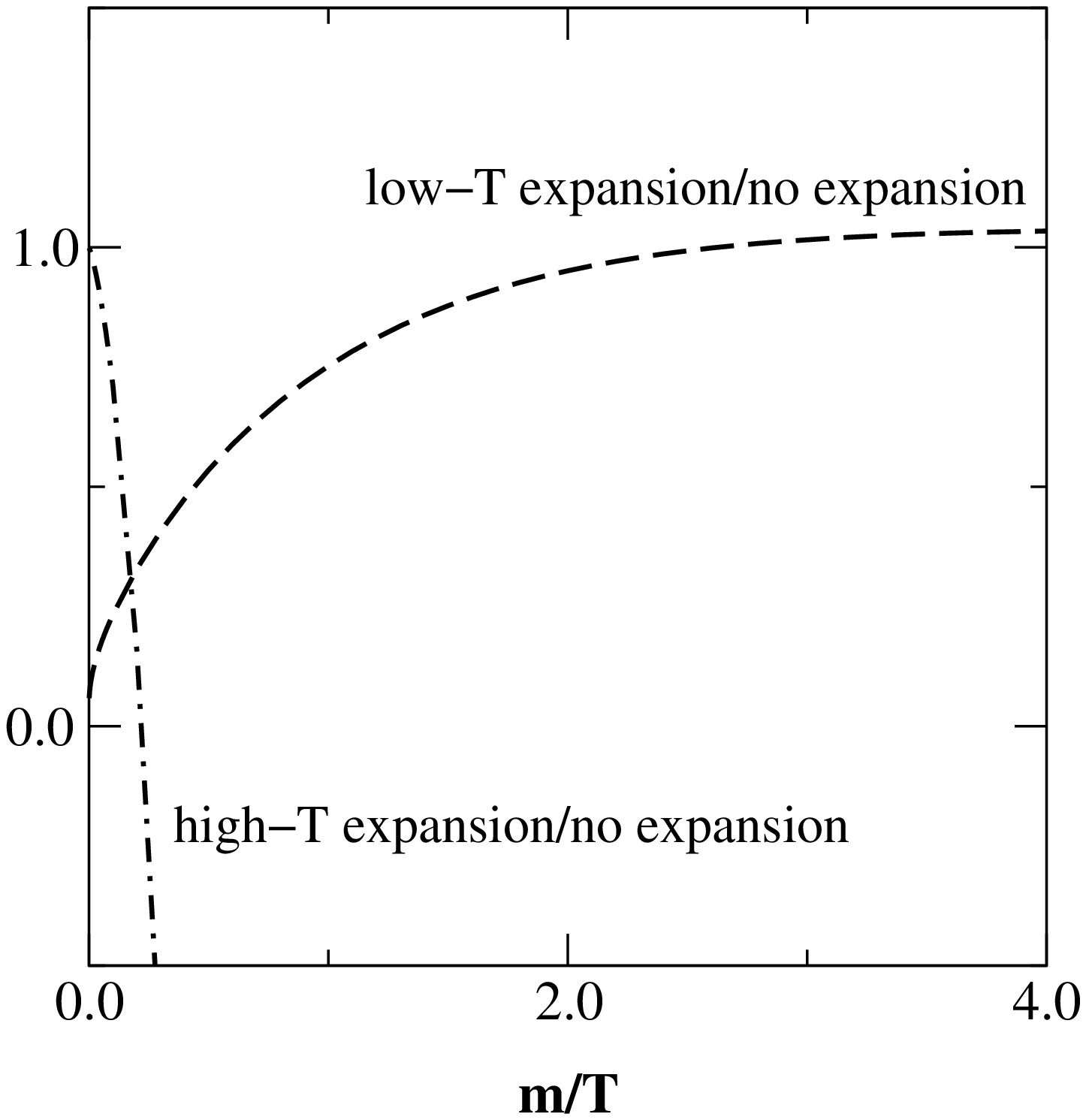}}


\caption[a]{Left: The function ${\cal H} \equiv H_\rmi{finite}$ 
in \eq\nr{Hfin}, 
along with the high-$T$ and low-$T$ limits in 
\eq\nr{Hfinlow}. Right: The ratio of the high-$T$ and low-$T$ limits
to the non-expanded result.}  
\la{fig:hnum}
\end{figure}

\noindent
The function ${\cal H}(m/T)$, numerically very easily evaluated, 
is plotted in \fig\ref{fig:hnum}, together with a comparison with 
the limiting values 
\ba
{\cal H}\Bigl(\frac{m}{T}\Bigr) & \stackrel{m\ll T}{=} & 
-\fr12 \biggl(\ln \frac{2 m}{T}-\fr13 + \gamma_E -
\frac{\zeta'(2)}{\zeta(2)} \biggr), \qquad 
\stackrel{m\gg T}{=} 
\frac{1}{2\pi} e^{-2m/T}. \la{Hfinlow}
\ea

\end{document}